\documentclass[aps,prl,twocolumn,superscriptaddress,nofootinbib,showpacs,letterpaper]{revtex4}

\usepackage{showlabels}
\usepackage{verbatim}
\usepackage[usenames,dvipsnames]{color}
\usepackage[normalem]{ulem}
\usepackage[pdftex]{graphicx}
\usepackage[english]{babel}
\usepackage{amsmath,amssymb}
\usepackage[colorlinks=true,citecolor=blue]{hyperref}
\usepackage{bm}
\usepackage{braket}
\usepackage{mathrsfs}

\begin{document}

\newcommand{\LL}{\mathcal{L}}
\newcommand{\DD}{\mathcal{D}}
\def\p{\bar{\rho}}

\title{The Quasinormal Modes of Weakly Charged Kerr-Newman Spacetimes}
\author{Zachary Mark}
\affiliation{Theoretical Astrophysics 350-17, California Institute of Technology, Pasadena, California 91125, USA}
\affiliation{Department of Physics and Astronomy, Oberlin College, Oberlin, Ohio, 44074, USA}

\author{Huan Yang}
\affiliation{Perimeter Institute for Theoretical Physics, Waterloo, Ontario, N2L2Y5, Canada}
\affiliation{Institute for Quantum Computing, University of Waterloo, Waterloo, Ontario N2L3G1, Canada}

\author{Aaron Zimmerman}
\affiliation{Canadian Institute for Theoretical Astrophysics, University of Toronto, Toronto, Ontario, M5S3H8, Canada}

\author{Yanbei Chen}
\affiliation{Theoretical Astrophysics 350-17, California Institute of Technology, Pasadena, California 91125, USA}

\begin{abstract}
The resonant mode spectrum of the Kerr-Newman spacetime is presently unknown.
These modes, called the quasinormal modes, play a central role in determining the stability of Kerr-Newman black holes and their response to perturbations.
We present a new formalism, generalized from time-independent perturbation theory in quantum mechanics, for calculating the quasinormal mode frequencies of weakly charged Kerr-Newman spacetimes of arbitrary spin. 
Our method makes use of an original technique for applying perturbation theory to zeroth-order solutions that are not square-integrable, and it can be applied to other problems in theoretical physics. 
The new formalism reveals no unstable modes, which together with previous results in the slow-rotation limit strongly indicates the modal stability of the Kerr-Newman spacetime. 
Our techniques and results are of interest in the areas of holographic duality, foundational problems in General Relativity, and possibly in astrophysical systems.
\end{abstract}


\maketitle

{\it Introduction.---} 
The resonant modes of a perturbed black hole spacetime are called quasinormal modes (QNMs) \cite{Berti2009}. They are found by solving an eigenvalue problem, similar to the type encountered in quantum mechanics, that arises from linearizing the Einstein-Maxwell (EM) equations about a stationary black hole background. The most general, stationary black hole solution in EM theory is the Kerr-Newman (KN) spacetime, which possesses a mass $M$, a specific angular momentum $a$, and an electric charge $Q$. The calculation of the QNM frequency spectrum of the perturbed KN spacetime is a major unsolved problem in General Relativity~\cite{ChandraBook}.

The KN QNM spectrum is a key component in a variety of problems involving charged black holes. Astrophysical black holes may temporarily acquire charge during compact binary mergers, as there could be nonzero charge distribution in their surrounding plasma~\cite{Alic2012,Lehner2012}, although in the stationary limit charges tend to be neutralized due to vacuum polarization~\cite{Gibbons75}. 
Computing the QNM frequencies is the first step in addressing the stability of the KN spacetime to perturbations, which remains an open question.
Knowledge of the KN QNMs would also assist efforts to determine whether the self force acts as a ``cosmic censor'' that prevents a KN black hole from being overcharged ($Q>Q_\text{max}=\sqrt{M^2-a^2}$) when a charged particle crosses the horizon~\cite{ZimmermanPoisson}. To determine the self force, the joint evolution of gravitational and electromagnetic fields induced by the motion of the charged particle 
must be evaluated consistently, a difficulty which may be amenable to the techniques we use here.
Studies of KN QNMs may be also relevant for string theory and holographic dualities~\cite{Berti2009}. In particular, according to the AdS/CFT correspondence~\cite{Maldacena:1997re}, the QNM spectrum of the bulk spacetime coincides with poles of the Green function of the boundary gauge theory. Extension of this work to KN-AdS~\cite{Caldarelli:1999xj} black holes can help to understand charged conformal fields on the boundary.

Current techniques to calculate QNM frequencies, such as Leaver's continued fraction method \cite{Leaver1985}, numerical shooting methods \cite{PressTeukolsky1973}, and newer techniques~\cite{Pani2013PHYSD,Pani:2013pma} require that the linearized EM equations separate.
The study of perturbations to both the charged, non-rotating Reissner-Nordstr\"{o}m (RN) black hole, and the rotating Kerr black hole can be reduced to the study of separable, decoupled wave equations, known as the Zerilli-Moncrief~\cite{Zerilli1974,Moncrief1974odd,Moncrief1974even} and Teukolsky equations~\cite{Teukolsky1973,PressTeukolsky1973,TeukolskyPress1974}, respectively. 

The problem of arriving at separable, decoupled equations governing gravitational and electromagnetic perturbations to the charged, rotating KN black hole is considerably harder. There is not enough symmetry to facilitate separation a priori, and the background electric field introduces interactions between the gravitational and electromagnetic perturbations which make decoupling nontrivial. 
Pani, Berti, and Gualtieri~\cite{Pani2013PRl,Pani2013PHYSD} dealt with these issues by working in the slow-rotation limit, where they found that the linearized KN equations separate and can be reduced to a pair of coupled ordinary differential equations. 
They were able to extract the QNM frequency spectrum using a matrix-valued version of Leaver's continued fraction method and their analysis revealed no unstable modes. 
Dudley and Finley derived a wave equation (hereafter referred to as the DF equation) that is exact for scalar perturbations, but is a conceptually questionable approximation for gravitational and electromagnetic modes. 
Berti and Kokkotas \cite{BertiKokkotas2005} conjectured that the DF equation is accurate for weakly-charged KN spacetimes by comparing its predictions for RN black holes to those from the Zerilli-Moncrief equation. 

In this paper we provide a new formalism, which we refer to as the eigenvalue perturbation (EVP) method, that is accurate to first order in $q\equiv Q^2/M^2$ and can be potentially extended to arbitrarily high orders in $q$. Our results show that the DF equation does not predict QNM frequencies that are accurate to first order in $q$, except for a special set of modes of rapidly rotating black holes. Our first order calculations in $q$ reveal no unstable modes, which, when complemented by the slow rotation study~\cite{Pani2013PRl,Pani2013PHYSD}, provides strong evidence for the linear stability of the KN spacetime. In addition, we provide the first analysis of the nearly extremal Kerr-Newman (NEKN) QNM frequencies in the rapidly rotating regime. The EVP method makes use of an original technique for applying perturbation theory to zeroth order solutions which are not square-integrable. This method can be (and recently has been \cite{Zimmerman:2014aha,Yang:2014zva,Yang:2014tla}) applied to other problems in theoretical physics.

{\it Problem.---} Following the derivation of the Teukolsky equation, we work in the Newman-Penrose (NP) formalism~\cite{NewmanPenrose1962,ChandraBook,StewartBook}, where the fundamental equations of General Relativity are projected onto a null tetrad. 
In the NP formalism, spin-weighted fields $\psi_s$ capture the information about the gravitational ($s=\pm 2$) and electromagnetic ($s =\pm 1$) perturbations. The spin-weighted fields are defined in terms of the Weyl scalars (for $s = \pm 1, \pm 2$) and are further discussed in the Appendix. 

We begin with the coupled equations for the scalars $\psi_1$ and $\psi_2$. Following Chandrasekhar \cite{ChandraBook}, we expand all NP quantities in frequency and azimuthal harmonics $e^{-i\omega t+im\phi}$, with harmonic indices implicit. Using similar steps as those used to derive the Teukolsky equation, we reduce the EM equations to two coupled equations governing the gravitational and electromagnetic degrees of freedom, which can schematically be written  
\begin{equation}
\begin{bmatrix}
\mathcal{H}_2 & q\delta \mathcal{H}_2 \\ \\
q \delta \mathcal{H}_1 &  \mathcal{H}_1 
\end{bmatrix}
\begin{bmatrix}
\psi_2 \\
\psi_1
\end{bmatrix}
=0.
\label{fun}
\end{equation}
The second order differential operators $\mathcal{H}_s$ and the $\delta \mathcal{H}_s$ operators contain derivatives in both $r$ and $\theta$. Their explicit form, as well as an outline of their derivation, is presented in~\cite{ChandraBook} and the Appendix. In all of the operators, the charge $Q$ only appears in the form $q = Q^2/M^2$, consistent with intuition that the frequencies cannot depend on the sign of the charge.
The $\delta \mathcal{H}_s$ operators are $\mathcal{O}(1)$ as $q \to 0$ [so the coupling terms are $\mathcal{O}(q)$] and Eq.~\eqref{fun} reduces to the $s=2$ and $s=1$ Teukolsky equations when $q \to 0$. The $\mathcal{H}_s$ operators differ from the unseparated, spin-weighted DF operator, which we denote by
$\mathcal{F}_s$, by an $\mathcal{O}(q)$ operator. This is why DF approximation fails to generate the correct order $\mathcal{O}(q)$ KN QNM frequencies. 

We are interested in the free oscillations of the KN spacetime, hence we supplement Eq.~\eqref{fun} with radiative boundary conditions; this means that we impose an ingoing boundary condition at the horizon and an outgoing boundary condition at spatial infinity. This turns Eq.~\eqref{fun} into an eigenvalue problem for the QNM frequency $\omega =\omega_R -i\omega_I$. A positive $\omega_I$ indicates a decaying, stable mode and a negative $\omega_I$ indicates a growing, unstable mode.
Similar coupled equations can be derived for the perturbations to the Weyl scalars $\psi_{-1}$ and $\psi_{-2}$; though we are not yet able to demonstrate it, we believe that as in Kerr, they yield the same QNM frequency spectrum. 

{\it Perturbative Formalism.---} We now calculate the QNM frequencies to $\mathcal{O}(q)$, which is the leading order correction to the frequency due the spacetime's charge. A solution to Eq.~\eqref{fun} consists of a frequency $\omega$, and a pair of spin-weighted fields $\psi_1$ and $\psi_2$ that satisfy the radiative boundary conditions. We expand our desired solution as a power series in $q$ around a QNM solution of Kerr,
\begin{align}
\psi_1 &= \psi_1^{(0)}+ q\psi_1^{(1)}+\mathcal{O}(q^2) \ , \nonumber\\ 
\psi_2 &= \psi_2^{(0)}+ q\psi_2^{(1)}+\mathcal{O}(q^2) \ , \nonumber\\ 
\omega &= \omega^{(0)}+ q\omega^{(1)}+\mathcal{O}(q^2) \ . \label{exp}
\end{align}
Either $\psi_1^{(0)}$ or $\psi_2^{(0)}$ are zero, since Eq.~\eqref{fun} decouples when $q=0$. Considering the gravitational perturbations first ($\psi_1^{(0)} = 0$) and plugging the expansions~\eqref{exp} into Eq.~\eqref{fun}, it is clear that the coupling terms enter at $\mathcal{O}(q^2)$ and can be neglected in our analysis. Expanding the $\mathcal{H}$ operators as a series in $q$, we obtain two decoupled equations for $\psi_2^{(1)}$ and $\psi_1^{(1)}$,
\begin{align}
&\mathcal{H}_2 \psi_2^{(1)} + \frac{\partial \mathcal{H}_2}{\partial q} \psi_2^{(0)}+ \omega^{(1)}\frac{\partial \mathcal{H}_2}{\partial \omega} \psi_2^{(0)} =0 \label{qmain}\,, \\
&\delta \mathcal{H}_1 \psi_2^{(0)}+ \mathcal{H}_1 \psi_1^{(1)}=0\,.
\end{align}
In the above expressions, we evaluate all operators at $q=0$ and $\omega = \omega^{(0)}$.

Equations of the form of Eq.~\eqref{qmain} are often encountered in quantum mechanics when we wish to calculate the corrections to the energy levels due to a small perturbing Hamiltonian. 
If we imagine that we likewise can define a finite product $\Braket{|}$ that makes the Teukolsky operator self-adjoint, meaning that $\Braket{\psi_s^{(0)} | \mathcal{H}_s \psi_s^{(1)} } = \Braket{\mathcal{H}_s \psi_s^{(0)} | \psi_s^{(1)} }= 0$ (with $\mathcal{H}_s$ again evaluated at $q =0$ and $\omega=\omega^{(0)}$),
we can derive a formula for $\omega^{(1)}$ by acting $\Bra{\psi_2^{(0)}}$ on both sides of  Eq.~\eqref{qmain}:
\begin{align} 
\omega^{(1)}=-\left .\Braket{\psi_2^{(0)}|\frac{\partial \mathcal{H}_2 }{\partial q}\psi_2^{(0)}} \right /\Braket{\psi_2^{(0)}|\frac{\partial \mathcal{H}_2 }{\partial \omega}\psi_2^{(0)}}\,.
\label{freqform}
\end{align}
An identical expression for the frequency correction to the electromagnetic QNM frequencies can be obtained from the same analysis with the $s=2$ subscript replaced by $s=1$. 
However, the QNM wave functions are not square-integrable along the real $r$-axis, since they diverge at the horizon and spatial infinity.
To derive a finite product, we observe that the outgoing boundary condition implies that the Teukolsky wave function $\psi_2^{(0)}$ exponentially decays as $r \to + i \infty$. By analytically continuing the QNM wave functions into the complex $r$-plane, we can define a finite product on two functions with the asymptotic behavior of Kerr QNM's:
\begin{align}
\Braket{\chi | \phi} \equiv \int \limits_\mathscr{C}(r-r_+)^s(r-r_-)^s dr \int \limits_0^\pi  \chi \phi \sin\theta d\theta\,,
\label{prod}
\end{align}
where $\mathscr{C}$ is a contour that starts to the right of $r_+$ at positive imaginary infinity, runs down parallel to the imaginary axis, encircles the point $r_+$ and returns to positive imaginary infinity, this time on the left of $r_+$. One might expect this contour integral to be zero by Cauchy's integral theorem; however, the functions in Eq.~\eqref{qmain} are not analytic in the the enclosed region. This is because the radial Teukolsky function has a branch point at $r_+$, and we use a branch cut that runs parallel to the imaginary axis emanating from $r_+$. The weights $(r-r_+)^s(r-r_-)^s$ and $\sin\theta$ are chosen to make the Teukolsky operator self-adjoint.

{\it Numerical Calculations.---} The spin-$s$ QNMs of a Kerr black hole are indexed by spheroidal harmonic indices $\ell$ and $m$, and an overtone number $n$. For a given $s$, $a$, $\ell$ and $m$, the least damped QNM is assigned $n=0$ (at least when there is no mode branching, see \cite{Yang2012b}).
We label the frequency corrections $\omega^{(1)}$ with the same indices as the corresponding background Kerr frequency $\omega^{(0)}$, grouping them as $\ell m n$. We only discuss the modes with $m \geq 0$ because of the symmetry $\omega(a,m) = \omega(-a,-m)$.

We explore the weakly charged KN QNM frequency spectrum by numerically evaluating Eq.~\eqref{freqform} for $\omega^{(1)}$. We use Leaver's continued fraction method to calculate the Kerr QNM frequencies $\omega^{(0)}$ and a truncated version of Leaver's expansion~\cite{Leaver1985} to represent the Teukolsky wave function $\psi_s^{(0)}$.
We estimate the error in our method in two ways. First, we perform the numerical integration twice for each mode, the second time keeping more terms in the wave function expansions and continued fractions, and also more points in the angular integral. We find that the fractional difference $|\omega^{(1)}_\text{run 2}-\omega^{(1)}_\text{run 1}|/|\omega^{(1)}_\text{run 2}|$ is roughly $10^{-6}$. Next, we estimate the error by applying the EVP method to the DF equation (i.e. we replace $\mathcal{H}_s$ with $\mathcal{F}_s$).  
The ``true'' DF QNM frequencies $\omega$ can be obtained via Leaver's method \cite{BertiKokkotas2005}, allowing $\omega^{(1)}$ to be computed independently of the EVP method via a numerical evaluation of $(\omega-\omega^{(0)})/q$ as $q \to 0$. 
In this way we find that the fractional error in $\omega^{(1)}$ is approximately $10^{-5}$.
The possible sources for these small errors are the truncation of the QNM wave functions, numerical imprecision in implementing the contour integral, and error in the root finding step of Leaver's method.

\begin{figure}[t]
\includegraphics[width=1.0\columnwidth]{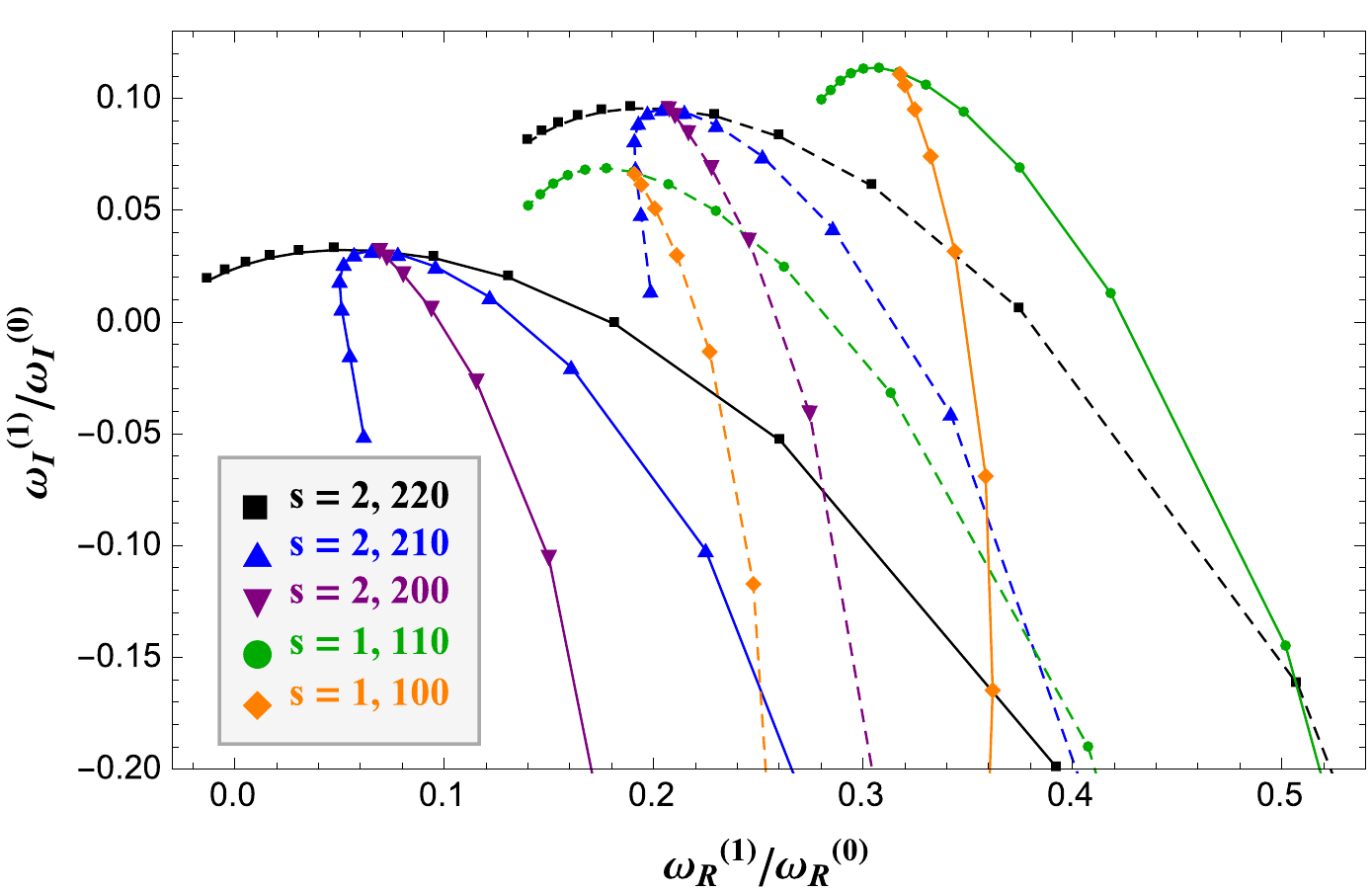}
\includegraphics[width=1.0\columnwidth]{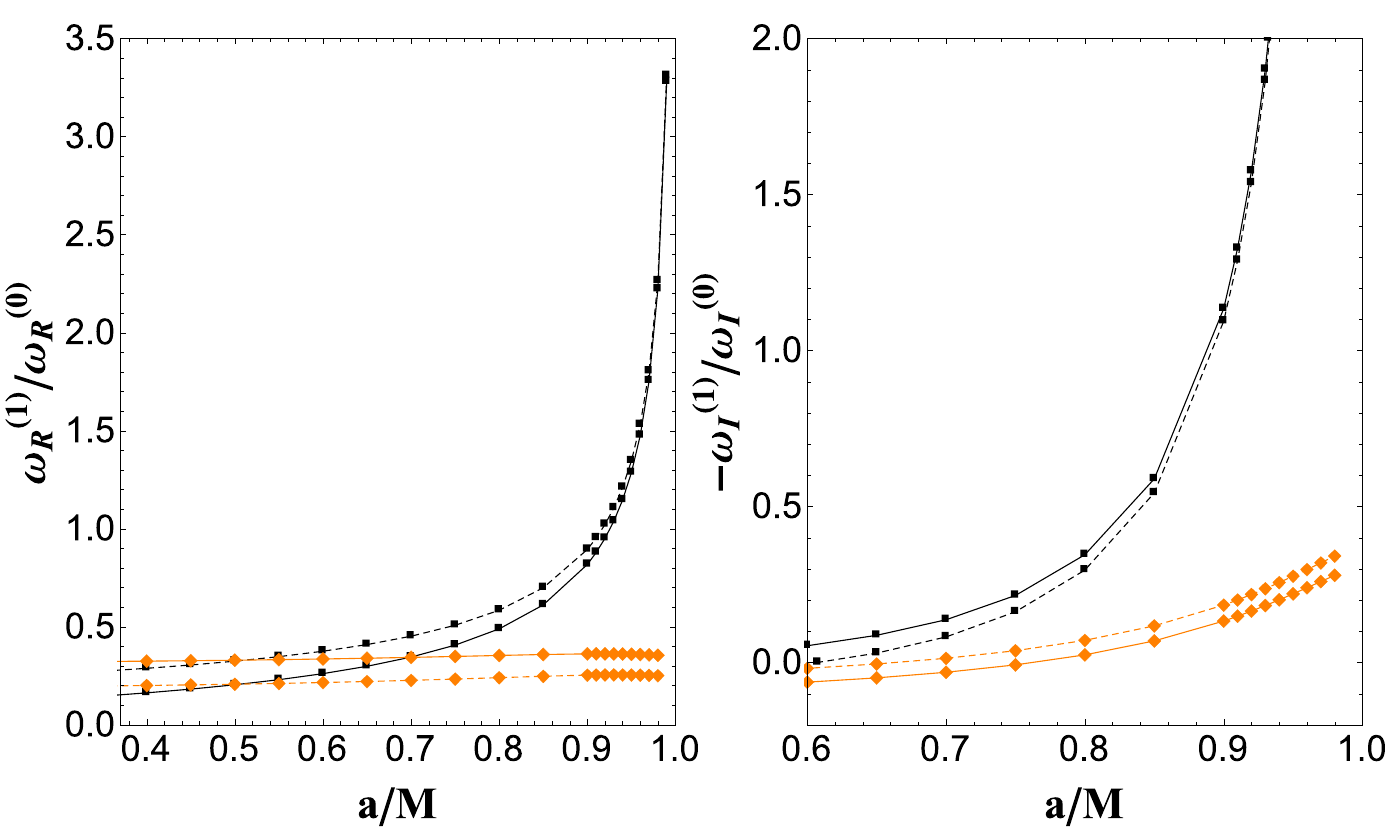}
\caption{The frequency corrections $\omega^{(1)}$ as predicted by the KN equations \eqref{fun} (solid lines) and by the DF equation (dashed lines). Top panel: scaled frequency corrections $\omega^{(1)}_R/\omega^{(0)}_R+i\omega^{(1)}_I/\omega^{(0)}_I$ as a function of $a/M$. Only the modes with $m \geq 0$ are plotted, and each subsequent data point increases by $0.15$ in $a/M$ (left to right), beginning with $a/M = -0.95$ for the $m=1,2$ modes and with $a/M=0$ for the $m=0$ modes. Bottom panels: The $s=2, 220$ and $s=1, 100$ QNM frequencies plotted versus $a/M$ in the rapidly-rotating regime.}
\label{fig:Corrections}
\end{figure}

In the top panel of Fig.~\ref{fig:Corrections}, we parametrically plot $\omega^{(1)}_R/\omega^{(0)}_R+i\omega^{(1)}_I/\omega^{(0)}_I$ in the complex plane as a function of $a/M$, for eight low-$\ell$ modes.
We show both the values of $\omega^{(1)}$, both as computed from Eq.~\eqref{freqform} (solid lines) and as predicted by the corresponding EVP analysis for DF equation (dashed lines).
We observe that in general there is a significant difference between the DF frequency corrections and the KN frequency corrections. The bottom panel of Fig.~\ref{fig:Corrections} focuses on the frequency corrections for rapidly-rotating black holes.
We plot $\omega^{(1)}_R/\omega^{(0)}_R$ and $\omega^{(1)}_I/\omega^{(0)}_I$ versus $a/M$ for large values of $a/M$. Notice that as $a \to M$, the DF equation predicts an increasingly accurate frequency correction $\omega^{(1)}$ for the $s=2$, $220$ mode, but not for the $s=1$, $100$ mode. 
We only plot two modes for clarity, but we also found that the DF equation becomes increasingly accurate as  $a \to M$ for the $s=1$, $110$ mode, but not for the $s=2$, $210$ or the the $s=2$, $200$ modes.

Using  Eq.~\eqref{freqform}, we can understand this phenomenon analytically. In the nearly extremal Kerr spacetime, there are two branches of QNMs \cite{Yang2012b, Yang:2013uba}; the Zero Damping Modes (ZDMs), which have zero decay in the extremal limit $a \to M$~\cite{Detweiler1980,Hod2008a}, and the Damped Modes (DMs), which retain a finite decay in this limit. The $s=2$, $220$ mode and the $s = 1$, $110$ mode are both ZDMs, while the $s=2$, $210$; $s=2$, $200$; and the $s=1$, $110$ modes are all DMs. By expanding the Teukolsky equation in powers of $\epsilon \equiv 1-a/M$, one can show that near the horizon ($r-r_+<\sqrt{\epsilon}$), the Kerr ZDMs depend on $\epsilon$ only through
the conformal variable $x \equiv (r-1)/\sqrt{\epsilon}$~\cite{TeukolskyPress1974,Yang:2013uba}, 
while DMs do not vary much with $\epsilon$ in the $\epsilon \rightarrow 0$ limit. Further, when analytically continued onto the contour $\mathscr{C}$, the ZDM wave function is concentrated in the near horizon region, allowing the integral $\eqref{prod}$ to be performed only over the near horizon region $x \ll 1$. 
Thus, we can figure out how the different terms in the formula for $\omega^{(1)}$ scale with $\epsilon$, if we write $\mathcal{F}_s$ and $\mathcal{H}_s$ in terms of the variable $x$ and then pick off the leading order $\epsilon$-dependence. The scalings are
\begin{align}
\frac{\partial \mathcal{F}_s}{\partial q}=\mathcal{O}(\epsilon^{-1}),\;\; \frac{\partial (\mathcal{H}_s-\mathcal{F}_s)}{\partial q} = \mathcal{O}(1) , \; \; 
 \frac{\partial \mathcal{H}_s}{\partial \omega} = \mathcal{O}(\epsilon^{-1/2}) \,.\label{scalings}
\end{align}
The DF equation predicts increasingly accurate frequency corrections as $\epsilon \to 0$ for modes which correspond to Kerr ZDMs because the term that it neglects in Eq.~\eqref{freqform} is $\mathcal{O}(\sqrt\epsilon)$, which is of subleading order.

\begin{figure}[t]
\includegraphics[width=1.0\columnwidth]{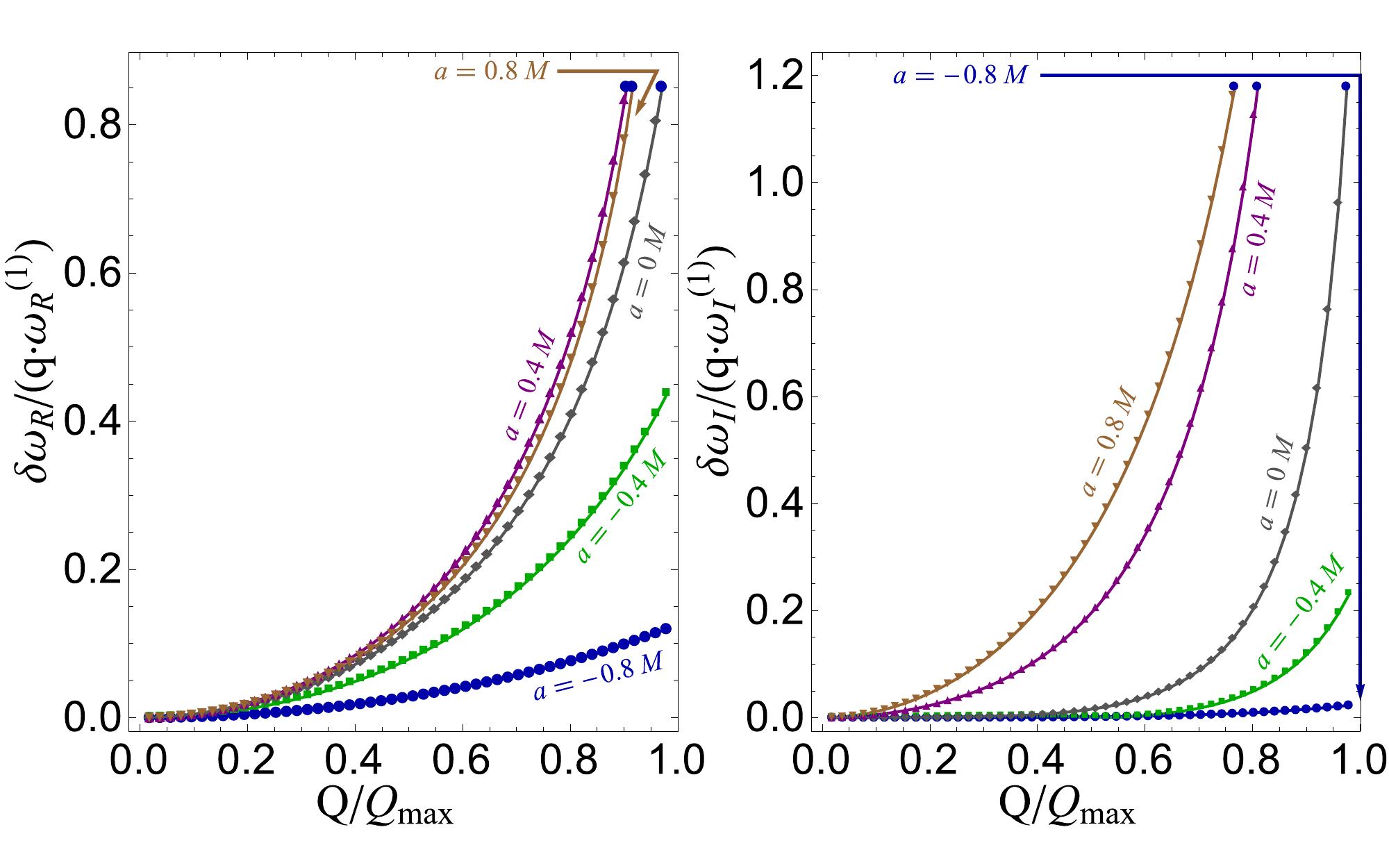}
\caption{Estimate of the size of higher order corrections in $q$, based on the EVP method applied to the DF equation. The residual error in the first order analysis is $\delta \omega = \omega -\omega^{(0)} -q\omega^{(1)}$, where $\omega$ is the true DF frequency calculated using Leaver's method. }
\label{fig:LimFirst}
\end{figure}

If we assume that our first order analysis in $q$ is accurate all the way up to $q_\text{max}$, none of the eight modes that we consider become unstable before they reach extremality. To estimate how large $Q$ can get before higher order contributions (in $q$) become important, we use the EVP method to calculate the leading order correction $\omega^{(1)}$ to the QNM frequencies of the DF equation.
We then calculate the residual error in the first order analysis $\delta \omega = \omega -\omega^{(0)} -q\omega^{(1)}$, where $\omega$ is the DF frequency calculated using Leaver's method, and compare it to $q\omega^{(1)}$. Figure \ref{fig:LimFirst} plots the comparison versus $Q/Q_\text{max}$ for the $s=2$, $220$ mode and selected values of $a$. We see that the importance of the higher order contributions varies greatly with $a$. Figure \ref{fig:LimFirst} also reveals that for most modes the first order analysis begins to fail long before $Q =Q_\text{max}$, indicating that going beyond linear analysis is likely necessary for NEKN QNMs.
However, there are some modes, such as the $a=-0.8\, M$, $s=2$, 220 mode, where the first order analysis is reasonably accurate, even when $Q = Q_\text{max}$. 

{\it Nearly Extremal Kerr-Newman.---} We now examine the modal stability of the weakly charged NEKN spacetime ($q < q_\text{max} =  2\epsilon -\epsilon^2 \ll 1$), where we have 
\begin{align}
\omega(\epsilon, q)=\omega^{(0)}(\epsilon)+\omega^{(1)}(\epsilon) q+ \omega^{(2)}(\epsilon) q^2+ \dots \,. \label{rapqexp}
\end{align}
Numerical searches and nearly extremal expansions \cite{ZimmermanDF} reveal that the extremal DF equation predicts marginally stable modes (ZDMs) for any value of $Q$, while previous work has not found ZDMs in the RN spacetime~\cite{Onozawa:1995vu, Andersson1996}. 
This tension can be resolved by working with the true KN perturbation equations. Suppose that $\omega^{(0)}$ is a Kerr ZDM. A nearly extremal analysis~\cite{Yang2012b} shows that $\omega_I^{(0)}=\mathcal{O}(\sqrt \epsilon)$. 
Substituting the scalings \eqref{scalings} into Eq.~\eqref{freqform}, reveals that $\omega^{(1)}=\mathcal{O}(\epsilon^{-1/2})$. The total charge correction $q \omega^{(1)}$ is in fact the same order as $\omega_I^{(0)}$ since $q_\text{max}\omega^{(1)}(\epsilon) =\mathcal{O}(\sqrt \epsilon)$, and may lead to a growing mode. 
Given our intuition from Fig.~\ref{fig:LimFirst}, confirming the existence of such a mode would require knowledge of the higher order charge corrections $q^j \omega^{(j)}$, which may also scale as $\mathcal{O}(\sqrt \epsilon)$. The stability of NEKN black holes and the possible existence of ZDMs remains an important open question which will be the subject of future investigation.
 
{\it Future Work.---} 
Our analysis provides the first calculation of the KN QNM frequencies for black holes with rapid rotation, and opens many avenues for the application of these results. A clear next step is to extend the analysis by computing the $\mathcal{O}(q)$ corrections to the wave functions, deepening our understanding of the coupling between the gravitational and electromagnetic field, and the $\mathcal{O}(q^2)$ frequency corrections, providing a better estimate of the error in the $\mathcal{O}(q)$ frequencies.

Finally, our analysis of NEKN black holes raises the question of whether ZDMs exist for nearly extremal black holes with arbitrary charge, which can be addressed with a more complete NEKN analysis. 
This would be complemented by a WKB analysis of the coupled equations~\eqref{fun}, would give further insights into the KN QNMs, the existence of ZDMs and DMs in the nearly extremal limit, and the possible geometric correspondence of the QNMs with geodesics~\cite{Mashhoon1985,Yang2012a}.

{\it Acknowledgement---} ZM would like to thank his undergraduate advisor Rob Owen for extensive discussions, both pedagogical and technical, and for looking over drafts of his thesis, which covers some of the same material. The authors also thank Luis Lehner and Yiqiu Ma for conversations about perturbations of KN black holes. AZ, HY, and YC were supported by NSF Grant PHY-1068881, CAREER Grant 0956189, and the David and Barbara Groce Startup Fund at Caltech. ZM was supported by the LIGO SURF program at Caltech. This research was supported in part by the Perimeter Institute for Theoretical Physics. Research at the Perimeter Institute is supported by the Government of Canada through Industry Canada and by the Province of Ontario through the Ministry of Research and Innovation.

\onecolumngrid
\appendix

\section{The Perturbed Kerr-Newman Spacetime in the Newman-Penrose Formalism}

The Newman-Penrose (NP) formalism \cite{ChandraBook, StewartBook, NewmanPenrose1962} is a null-tetrad formulation of General Relativity that offers a simple way of describing spacetimes with one or more sheer-free congruences of null geodesics. Like the Kerr spacetime, the Kerr-Newman (KN) spacetime possesses two such congruences and as a consequence when an appropriate tetrad is chosen the spin coefficients $\kappa$, $\sigma$, $\lambda$, and $\nu$, the Weyl scalars $\Psi_0$, $\Psi_1$, $\Psi_3$, and $\Psi_4$, as well as the Maxwell scalars $\phi_0$ and $\phi_2$ vanish. Hence, in the perturbed KN spacetime these quantities are all of perturbative order and the linearized NP field equations are greatly simplified.  

Despite the similarity in the NP descriptions of the Kerr and KN spacetimes, the perturbed KN spacetime is far more complicated than the perturbed Kerr spacetime. 
While gravitational perturbations and electromagnetic perturbations can be independently excited in the Kerr spacetime, they are necessarily intricately intertwined in the KN spacetime. 
Thus, the perturbed KN spacetime contains two families of perturbations; one family becomes the gravitational perturbations in the $Q \to 0$ limit, while the other family becomes the electromagnetic perturbations. 

In the perturbed KN spacetime (using the appropriate tetrad), the Weyl scalars $\Psi_0$ and $\Psi_4$ are gauge invariant (under infinitesimal tetrad transformations) and they describe gravitational waves near the horizon and near null infinity, respectively.  The rest of the Weyl scalars and Maxwell scalars are not gauge invariant, as is also true in the perturbed Kerr spacetime. There are two convenient gauges to consider. In the perturbed Kerr spacetime, the standard choice is to set $\Psi_1$ and $\Psi_3$ equal to zero by the appropriate infinitesimal tetrad transformation. This means that $\phi_0$ and $\phi_2$  are nonzero and they  describe electromagnetic radiation near the horizon and near infinity, respectively. In the Kerr spacetime, $\phi_0$ and $\phi_2$  can be perturbed independently from $\Psi_0$ and $\Psi_4$, while in the KN spacetime, all four perturbations must be excited simultaneously. Alternatively, one can use the ``phantom gauge'' \cite{ChandraBook} and set $\phi_0$ and $\phi_2$ equal to zero. The Weyl scalars $\Psi_1$ and $\Psi_3$ then contain the information describing the electromagnetic field. One way of understanding this is that knowledge of $\Psi_1$ and $\Psi_3$ is necessary to recover  $\phi_0$ and $\phi_2$ in the standard gauge choice. For our work, we adopt the phantom gauge because the linearized NP field equations appear to be simplest in the phantom gauge. 

We now use the least coupled, linearized NP equations to derive a pair of coupled equations governing $\Psi_0$ and $\Psi_1$ (or $\Psi_3$ and $\Psi_4$). The are many ways of obtaining such equations, but the equations that we arrive at reduce to the Teukolsky equation in the $Q \to 0$ limit and clarify the relationship of the DF equation to the ``true'' KN linearized equations.

We follow Chandrasekhar, using the Kinnersley tetrad and Boyer-Linquist coordinates, and we expand all NP quantities in frequency and azimuthal harmonics  $e^{-i\omega t+im\phi}$. We also adopt Chandrasekhar's notation, defining the following operators which arise when the directional derivative operators $\mathbf{D}$, $\mathbf{\Delta}$, $\boldsymbol{\delta}$, and $\boldsymbol{\delta}^*$ act on functions that are expanded in frequency and azimuthal harmonics:
\begin{align}
&\DD_j \equiv \partial_r+\frac{iK}{\Delta}+2j\frac{(r-M)}{\Delta},& && &\DD_j^\dagger \equiv \partial_r-\frac{iK}{\Delta}+2j\frac{(r-M)}{\Delta}& \nonumber
\\ &\LL_j \equiv \partial_\theta+\hat{Q}+j\cot\theta,& && &\LL_j^\dagger \equiv \partial_\theta-\hat{Q}+j\cot\theta,& \nonumber
\\ &K=-(r^2+a^2)\omega+am,& &\Delta = r^2 -2Mr+a^2+q,& &\hat{Q}=-a\omega\sin\theta+\frac{m}{\sin\theta}.& \nonumber
\end{align}
We define spin weighted fields that capture the gravitational and electromagnetic degrees of freedom.
\begin{align}
\psi_{-2} =\p^{*4}\Psi_4, & & \psi_{-1}=\frac{\p^{*3}\Psi_3}{\sqrt{2}} , & & \psi_1=\sqrt{2} \p^*\Psi_1, & & \psi_2=\Psi_0,
\end{align}
where $\bar{\rho}=r+ia\cos\theta$, $\bar{\rho}=r-ia\cos\theta$, and the prefactors  are necessary to separate the Teukolsky equation in Kerr.
Further, we define scaled versions of the spin coefficients
\begin{align}
k=\frac{\kappa}{\sqrt{2}\p^{*2}}, & & s=\frac{\p\tilde\sigma}{\p^{*2}}, & & \ell =\frac{\p^*\lambda}{2}, & &n=\frac{\rho^2\nu}{\sqrt{2}},
\end{align}
where $\rho^2=\bar \rho \bar \rho^*$.

We begin by linearizing the NP equations Chandrasekhar Ch. 1, (321a) (a Bianchi Identity), (321e) (a Bianchi Identity),  (310b) (a Ricci Identity); Chandrasekhar Ch 11, (136) (a manipulated version of several Maxwell equations); and their GHP transforms \cite{GHP}. The first set of four equations, from which we derive a pair of coupled equations governing $\psi_1$ and $\psi_2$, are expressed in Boyer-Lindquist coordinates as:
\begin{align}
&\left[\LL_2-\frac{3ia\sin\theta}{\p^*} \right]\psi_2 -\left[\DD_0+\frac{3}{\p^*}\right]\psi_1=-2k\left[3\left(M-\frac{Q^2}{\p}\right)+\frac{Q^2\p^*}{\p^2}\right] \,,\label{j1}
\\& \Delta\left[\DD_2^\dagger-\frac{3}{\p^*}\right]\psi_2+\left[\LL_{-1}^\dagger+\frac{3ia\sin\theta}{\p^*}\right]\psi_1=2s\left[3(M-\frac{Q^2}{\p})-\frac{Q^2\p^*}{\p^2}\right] \,, \label{j2}
\\ &\left[\DD_0+\frac{3}{\p^*}\right]s-\left[\LL_{-1}^\dagger+\frac{3ia\sin\theta}{\p^*}\right]k=\frac{\p}{\p^{*2}}\psi_2 \,, \label{j3}
\\ &\Delta\left[\DD_2^\dagger-\frac{3}{\p^*}\right]k+\left[\LL_2-\frac{3ia\sin\theta}{\p^*}\right]s=2\frac{\p}{\p^{*2}}\psi_1 \,. \label{j4}
\end{align}
The GHP transformed versions of these particular equations, from which we derive a pair of coupled equations governing  $\psi_{-1}$ and $\psi_{-2}$, are obtained by replacing 
\begin{align}
&\psi_1 \to -\psi_{-1},& &k \to -n,& & \LL_{-1}^\dagger \to \LL_{-1},& &\left[ \DD_0 +\frac{3}{\bar \rho^*}\right] \to \Delta \left[ \DD_{-1}^\dagger +\frac{3}{\bar \rho^*}\right],& \nonumber
\\ & \psi_2 \to \psi_{-2},& & s \to \ell, & &\LL_2 \to \LL_2^\dagger,& & \Delta\left[\DD_2^\dagger -\frac{3}{\bar \rho^*}\right] \to \left[\DD_0 -\frac{3}{\bar \rho^*}\right].&
\end{align}
As we are trying to get our equations in a form similar to that of the Teukolsky equation, we apply the same manipulations to the KN perturbation equations that decouple the Teukolsky equation. We obtain equations for $\psi_{-2}$, $\psi_{-1}$, $\psi_1$, and $\psi_2$ coupled to only the spin coefficients (in the $Q=0$ case, there is no coupling to the spin coefficients and these equations are the Teukolsky equations) by making use of the commutation relation
\begin{align}
\left[\DD+\frac{c}{\p^*}\right]\left[\LL+\frac{iac\sin\theta}{\p^*}\right]=\left[\LL+\frac{iac\sin\theta}{\p^*}\right]\left[\DD+\frac{c}{\p^*}\right], \label{mcrel}
\end{align}
where $\DD$ represents any of the $\DD_j$ operators or $\DD^\dagger_j$ operators of any $j$, $\LL$ represents any of the $\LL_{j'}$ operators or $\LL^\dagger_{j'}$ operators, of any $j'$ ($j'$ does not have to equal $j$), and $c$ is any constant. The simplified equations are
\begin{align}
&\left(\Delta\DD_1\DD_2^\dagger+\LL_{-1}^\dagger\LL_2+6i\omega\p\right)\psi_2=-2Q^2\left[\LL_{-1}^\dagger\left(\frac{\p^*k}{\p^2}\right)+\DD_0\left(\frac{\p^*s}{\p^2}\right)\right], \label{p1}
\\ & \left(\Delta\DD_2^\dagger\DD_0+\LL_2\LL_{-1}^\dagger+6i\omega\p\right)\psi_1=2Q^2\left[\Delta\DD_2^\dagger\left(\frac{\p^*k}{\p^2}\right)-\LL_2\left(\frac{\p^*s}{\p^2}\right)\right], \label{p2}
\\& \left(\Delta\DD_1\DD_{-1}^\dagger+\LL_2^\dagger\LL_{-1}-6i\omega\p\right)\psi_{-1}= 2Q^2\left[\DD_0\left(\frac{\p^* n}{\p^2}\right)+\LL_2^\dagger\left(\frac{\p^*\ell}{\p^2}\right)\right], \label{p3}
\\& \left(\Delta\DD_{-1}^\dagger\DD_0+\LL_{-1}\LL_2^\dagger-6i\omega\p\right)\psi_{-2} = -2Q^2\left[-\LL_{-1}\left(\frac{\p^* n}{\p^2}\right)+\Delta\DD_{-1}^\dagger\left(\frac{\p^* \ell}{\p^2}\right)\right]. \label{p4}
\end{align}
We arrive at our desired equations by solving Eqs.~\eqref{j1}, \eqref{j2}, and their GHP transforms for $k$, $s$, $n$, and $\ell$ and inserting the resulting expressions into Eqs.~\eqref{p1}, \eqref{p2}, \eqref{p3}, and \eqref{p4}. The final equations are
\begin{equation}
\begin{bmatrix}
\mathcal{F}_{\pm2}+ q\mathcal{G}_{\pm2}  & q\delta \mathcal{H}_{\pm2} \\ \\
q \delta \mathcal{H}_{\pm1} &  \mathcal{F}_{\pm1} +q\mathcal{G}_{\pm1}
\end{bmatrix}
\begin{bmatrix}
\psi_{\pm2} \\
\psi_{\pm1}
\end{bmatrix}
=0.
\label{fun2}
\end{equation}
The $\mathcal{F}_s$, $\mathcal{G}_s$, and $\delta \mathcal{H}_s$ operators are defined in Table 1, where we have defined $\alpha_\pm \equiv \left[3(\p^2M-\p Q^2)\pm Q^2\p^*\right]^{-1}$ in order to simplify the expressions. The $\mathcal{F}_s$ operator is the spin $s$ DF operator, which becomes the Teukolsky operator in the $Q \to 0$ limit. The only $q$ dependence in $\mathcal{F}_s$ comes from $\Delta$, and the DF equation can be understood as the Teukolsky equation with modification $\Delta_\text{kerr} =r^2-2Mr+a^2 \to \Delta$. The $\mathcal{O}(1)$ operator $\delta\mathcal{H}_s$ introduces an $\mathcal{O}(q)$ coupling term into the perturbation equations. Finally, the operator $q\mathcal{G}_s \psi_s$ is the $\mathcal{O}(q)$ difference between the $\mathcal{H}_s$ term and the $\mathcal{F}_s$ term in Eq.~\eqref{fun2} alluded to in the body of the paper. The $\mathcal{F}_s$ and $\mathcal{G}_s$ operators are important to the $\mathcal{O}(q)$ analysis of the EVP method, while the  $\delta\mathcal{H}_s$ operator comes in at higher order and can be neglected.

\begin{table}
\begin{center}
    \begin{tabular}{ | c | c | c |  c|}
    \hline
    $s$ & $\mathcal{F}_s f$ & $\mathcal{G}_s f$ & $\delta\mathcal{H}_s f$ \\ \hline
  2 & $ \Delta \DD_1\DD_2^\dagger f +\LL_{-1}^\dagger\LL_2 f+6i\omega\bar\rho f$
 & $\LL_{-1}^\dagger \alpha_+3ia\sin\theta f-\LL_{-1}^\dagger \alpha_+\p^*\LL_2 f$ & $\LL_{-1}^\dagger\alpha_+\p^*\DD_0f+3\LL_{-1}^\dagger\alpha_+f $\\ & &$+\DD_0\alpha_-\Delta\p^*\DD_2^\dagger f-3\DD_0\alpha_- \Delta f$ & $+\DD_0\alpha_-\p^*\LL_{-1}^\dagger f+3\DD_0\alpha_-ia\sin\theta f$  \\ \hline
    1 & $ \Delta \DD_2^\dagger\DD_0 f +\LL_{2}\LL_{-1}^\dagger f+6i\omega\bar\rho f$ & $-\Delta\DD_2^\dagger\alpha_+\p^*\DD_0f-3\Delta\DD_2^\dagger\alpha_+f$ & $-3\Delta\DD_2^\dagger\alpha_+ia\sin\theta f+\Delta\DD_2^\dagger\alpha_+\p^*\LL_2f$ \\ & &$+\LL_2\alpha_-\p^*\LL_{-1}^\dagger f+3\LL_2\alpha_-ia\sin\theta f$ & $+\LL_2\alpha_-\Delta\p^*\DD_2^\dagger f-3\LL_2\alpha_-\Delta f $\\ \hline
    -1 & $\Delta\DD_1\DD_{-1}^\dagger f+\LL_2^\dagger\LL_{-1}f-6i\p\omega f$ & $- \DD_0 \alpha_+ \p^* \Delta \DD_{-1}^\dagger f -3 \DD_0 \alpha_+ \Delta f$ & $-\DD_0 \alpha_+ \p^* \LL_2^\dagger f +3 \DD_0 \alpha_+ ia \sin \theta f$ \\ & &$\LL_2^\dagger \alpha_- \p^* \LL_{-1} f +3 \LL_2^\dagger \alpha_- i a\sin\theta f$ &$-\LL_2^\dagger \alpha_- \p^* \DD_0 f +3 \LL_2^\dagger \alpha _- f$ \\ \hline
    -2 & $\Delta\DD_{-1}^\dagger\DD_0 f+\LL_{-1}\LL_2^\dagger f-6i\p\omega f$& $\Delta \DD_{-1}^\dagger \alpha_-\p^*\DD_0 f -3\Delta \DD_{-1}^\dagger \alpha_- f$ & $-\Delta \DD_{-1}^\dagger \alpha_- \p^* \LL_{-1} f -3 \Delta \DD_{-1}^\dagger \alpha_- i a \sin \theta f$ \\ &  &$- \LL_{-1}\alpha_+ \p^* \LL_2 ^\dagger f +3 \LL_{-1} \alpha_+ i a \sin \theta f$ & $-\LL_{-1} \alpha_+ \p^* \Delta \DD_{-1}^\dagger f -3\LL_{-1} \alpha_+ \Delta f$ \\ \hline
    \end{tabular}
\end{center}
\caption{The definitions of the $\mathcal{F}_s$, $\mathcal{G}_s$, and $\delta \mathcal{H}_s$ operators.}
\end{table}

\bibliography{Refs}

\begin{thebibliography}{34}
\expandafter\ifx\csname natexlab\endcsname\relax\def\natexlab#1{#1}\fi
\expandafter\ifx\csname bibnamefont\endcsname\relax
  \def\bibnamefont#1{#1}\fi
\expandafter\ifx\csname bibfnamefont\endcsname\relax
  \def\bibfnamefont#1{#1}\fi
\expandafter\ifx\csname citenamefont\endcsname\relax
  \def\citenamefont#1{#1}\fi
\expandafter\ifx\csname url\endcsname\relax
  \def\url#1{\texttt{#1}}\fi
\expandafter\ifx\csname urlprefix\endcsname\relax\def\urlprefix{URL }\fi
\providecommand{\bibinfo}[2]{#2}
\providecommand{\eprint}[2][]{\url{#2}}

\bibitem[{\citenamefont{{Berti} et~al.}(2009)\citenamefont{{Berti}, {Cardoso},
  and {Starinets}}}]{Berti2009}
\bibinfo{author}{\bibfnamefont{E.}~\bibnamefont{{Berti}}},
  \bibinfo{author}{\bibfnamefont{V.}~\bibnamefont{{Cardoso}}},
  \bibnamefont{and} \bibinfo{author}{\bibfnamefont{A.~O.}
  \bibnamefont{{Starinets}}}, \bibinfo{journal}{Classical and Quantum Gravity}
  \textbf{\bibinfo{volume}{26}}, \bibinfo{pages}{163001}
  (\bibinfo{year}{2009}), \eprint{0905.2975}.

\bibitem[{\citenamefont{Chandrasekhar}(1983)}]{ChandraBook}
\bibinfo{author}{\bibfnamefont{S.}~\bibnamefont{Chandrasekhar}},
  \emph{\bibinfo{title}{The Mathematical Theory of Black Holes}}
  (\bibinfo{publisher}{Clarendon Press. Oxford}, \bibinfo{year}{1983}).

\bibitem[{\citenamefont{Alic et~al.}(2012)\citenamefont{Alic, Moesta, Rezzolla,
  Zanotti, and Jaramilli}}]{Alic2012}
\bibinfo{author}{\bibfnamefont{D.}~\bibnamefont{Alic}},
  \bibinfo{author}{\bibfnamefont{P.}~\bibnamefont{Moesta}},
  \bibinfo{author}{\bibfnamefont{L.}~\bibnamefont{Rezzolla}},
  \bibinfo{author}{\bibfnamefont{O.}~\bibnamefont{Zanotti}}, \bibnamefont{and}
  \bibinfo{author}{\bibfnamefont{J.}~\bibnamefont{Jaramilli}},
  \bibinfo{journal}{\apj} \textbf{\bibinfo{volume}{754}}, \bibinfo{pages}{36}
  (\bibinfo{year}{2012}), \eprint{1204.2226}.

\bibitem[{\citenamefont{Lehner et~al.}(2012)\citenamefont{Lehner, Palenzuela,
  Liebling, Thompson, and Hanner}}]{Lehner2012}
\bibinfo{author}{\bibfnamefont{L.}~\bibnamefont{Lehner}},
  \bibinfo{author}{\bibfnamefont{C.}~\bibnamefont{Palenzuela}},
  \bibinfo{author}{\bibfnamefont{S.}~\bibnamefont{Liebling}},
  \bibinfo{author}{\bibfnamefont{C.}~\bibnamefont{Thompson}}, \bibnamefont{and}
  \bibinfo{author}{\bibfnamefont{C.}~\bibnamefont{Hanner}},
  \bibinfo{journal}{\prd} \textbf{\bibinfo{volume}{86}}, \bibinfo{eid}{104035}
  (\bibinfo{year}{2012}), \eprint{1112.2633}.

\bibitem[{\citenamefont{Gibbons}(245)}]{Gibbons75}
\bibinfo{author}{\bibfnamefont{G.~W.} \bibnamefont{Gibbons}},
  \bibinfo{journal}{Commun. math. Phys.} \textbf{\bibinfo{volume}{44}},
  \bibinfo{pages}{245} (\bibinfo{year}{245}).

\bibitem[{\citenamefont{Zimmerman and Poisson}(2014)}]{ZimmermanPoisson}
\bibinfo{author}{\bibfnamefont{P.}~\bibnamefont{Zimmerman}} \bibnamefont{and}
  \bibinfo{author}{\bibfnamefont{E.}~\bibnamefont{Poisson}}
  (\bibinfo{year}{2014}).

\bibitem[{\citenamefont{Maldacena}(1999)}]{Maldacena:1997re}
\bibinfo{author}{\bibfnamefont{J.~M.} \bibnamefont{Maldacena}},
  \bibinfo{journal}{Int.J.Theor.Phys.} \textbf{\bibinfo{volume}{38}},
  \bibinfo{pages}{1113} (\bibinfo{year}{1999}), \eprint{hep-th/9711200}.

\bibitem[{\citenamefont{Caldarelli et~al.}(2000)\citenamefont{Caldarelli,
  Cognola, and Klemm}}]{Caldarelli:1999xj}
\bibinfo{author}{\bibfnamefont{M.~M.} \bibnamefont{Caldarelli}},
  \bibinfo{author}{\bibfnamefont{G.}~\bibnamefont{Cognola}}, \bibnamefont{and}
  \bibinfo{author}{\bibfnamefont{D.}~\bibnamefont{Klemm}},
  \bibinfo{journal}{Class.Quant.Grav.} \textbf{\bibinfo{volume}{17}},
  \bibinfo{pages}{399} (\bibinfo{year}{2000}), \eprint{hep-th/9908022}.

\bibitem[{\citenamefont{Leaver}(1985)}]{Leaver1985}
\bibinfo{author}{\bibfnamefont{E.}~\bibnamefont{Leaver}},
  \bibinfo{journal}{Proc.Roy.Soc.Lond.} \textbf{\bibinfo{volume}{A402}},
  \bibinfo{pages}{285} (\bibinfo{year}{1985}).

\bibitem[{\citenamefont{{Press} and {Teukolsky}}(1973)}]{PressTeukolsky1973}
\bibinfo{author}{\bibfnamefont{W.~H.} \bibnamefont{{Press}}} \bibnamefont{and}
  \bibinfo{author}{\bibfnamefont{S.~A.} \bibnamefont{{Teukolsky}}},
  \bibinfo{journal}{\apj} \textbf{\bibinfo{volume}{185}}, \bibinfo{pages}{649}
  (\bibinfo{year}{1973}).

\bibitem[{\citenamefont{Pani et~al.}(2013{\natexlab{a}})\citenamefont{Pani,
  Berti, and Gualtieri}}]{Pani2013PHYSD}
\bibinfo{author}{\bibfnamefont{P.}~\bibnamefont{Pani}},
  \bibinfo{author}{\bibfnamefont{E.}~\bibnamefont{Berti}}, \bibnamefont{and}
  \bibinfo{author}{\bibfnamefont{L.}~\bibnamefont{Gualtieri}},
  \bibinfo{journal}{Phys. Rev. D} \textbf{\bibinfo{volume}{88}},
  \bibinfo{pages}{064048} (\bibinfo{year}{2013}{\natexlab{a}}),
  \urlprefix\url{http://link.aps.org/doi/10.1103/PhysRevD.88.064048}.

\bibitem[{\citenamefont{Pani}(2013)}]{Pani:2013pma}
\bibinfo{author}{\bibfnamefont{P.}~\bibnamefont{Pani}},
  \bibinfo{journal}{Int.J.Mod.Phys.} \textbf{\bibinfo{volume}{A28}},
  \bibinfo{pages}{1340018} (\bibinfo{year}{2013}), \eprint{1305.6759}.

\bibitem[{\citenamefont{Zerilli}(1974)}]{Zerilli1974}
\bibinfo{author}{\bibfnamefont{F.~J.} \bibnamefont{Zerilli}},
  \bibinfo{journal}{Phys. Rev. D} \textbf{\bibinfo{volume}{9}},
  \bibinfo{pages}{860} (\bibinfo{year}{1974}),
  \urlprefix\url{http://link.aps.org/doi/10.1103/PhysRevD.9.860}.

\bibitem[{\citenamefont{Moncrief}(1974{\natexlab{a}})}]{Moncrief1974odd}
\bibinfo{author}{\bibfnamefont{V.}~\bibnamefont{Moncrief}},
  \bibinfo{journal}{Phys. Rev. D} \textbf{\bibinfo{volume}{9}},
  \bibinfo{pages}{2707} (\bibinfo{year}{1974}{\natexlab{a}}),
  \urlprefix\url{http://link.aps.org/doi/10.1103/PhysRevD.9.2707}.

\bibitem[{\citenamefont{Moncrief}(1974{\natexlab{b}})}]{Moncrief1974even}
\bibinfo{author}{\bibfnamefont{V.}~\bibnamefont{Moncrief}},
  \bibinfo{journal}{Phys. Rev. D} \textbf{\bibinfo{volume}{10}},
  \bibinfo{pages}{1057} (\bibinfo{year}{1974}{\natexlab{b}}),
  \urlprefix\url{http://link.aps.org/doi/10.1103/PhysRevD.10.1057}.

\bibitem[{\citenamefont{{Teukolsky}}(1973)}]{Teukolsky1973}
\bibinfo{author}{\bibfnamefont{S.~A.} \bibnamefont{{Teukolsky}}},
  \bibinfo{journal}{\apj} \textbf{\bibinfo{volume}{185}}, \bibinfo{pages}{635}
  (\bibinfo{year}{1973}).

\bibitem[{\citenamefont{{Teukolsky} and {Press}}(1974)}]{TeukolskyPress1974}
\bibinfo{author}{\bibfnamefont{S.~A.} \bibnamefont{{Teukolsky}}}
  \bibnamefont{and} \bibinfo{author}{\bibfnamefont{W.~H.}
  \bibnamefont{{Press}}}, \bibinfo{journal}{\apj}
  \textbf{\bibinfo{volume}{193}}, \bibinfo{pages}{443} (\bibinfo{year}{1974}).

\bibitem[{\citenamefont{Pani et~al.}(2013{\natexlab{b}})\citenamefont{Pani,
  Berti, and Gualtieri}}]{Pani2013PRl}
\bibinfo{author}{\bibfnamefont{P.}~\bibnamefont{Pani}},
  \bibinfo{author}{\bibfnamefont{E.}~\bibnamefont{Berti}}, \bibnamefont{and}
  \bibinfo{author}{\bibfnamefont{L.}~\bibnamefont{Gualtieri}},
  \bibinfo{journal}{Phys. Rev. Lett.} \textbf{\bibinfo{volume}{110}},
  \bibinfo{pages}{241103} (\bibinfo{year}{2013}{\natexlab{b}}),
  \urlprefix\url{http://link.aps.org/doi/10.1103/PhysRevLett.110.241103}.

\bibitem[{\citenamefont{{Berti} and {Kokkotas}}(2005)}]{BertiKokkotas2005}
\bibinfo{author}{\bibfnamefont{E.}~\bibnamefont{{Berti}}} \bibnamefont{and}
  \bibinfo{author}{\bibfnamefont{K.~D.} \bibnamefont{{Kokkotas}}},
  \bibinfo{journal}{\prd} \textbf{\bibinfo{volume}{71}}, \bibinfo{eid}{124008}
  (\bibinfo{year}{2005}), \eprint{arXiv:gr-qc/0502065}.

\bibitem[{\citenamefont{Zimmerman et~al.}(2014)\citenamefont{Zimmerman, Yang,
  Mark, Chen, and Lehner}}]{Zimmerman:2014aha}
\bibinfo{author}{\bibfnamefont{A.}~\bibnamefont{Zimmerman}},
  \bibinfo{author}{\bibfnamefont{H.}~\bibnamefont{Yang}},
  \bibinfo{author}{\bibfnamefont{Z.}~\bibnamefont{Mark}},
  \bibinfo{author}{\bibfnamefont{Y.}~\bibnamefont{Chen}}, \bibnamefont{and}
  \bibinfo{author}{\bibfnamefont{L.}~\bibnamefont{Lehner}}
  (\bibinfo{year}{2014}), \eprint{arXiv:1406.4206}.

\bibitem[{\citenamefont{Yang and Zhang}(2014)}]{Yang:2014zva}
\bibinfo{author}{\bibfnamefont{H.}~\bibnamefont{Yang}} \bibnamefont{and}
  \bibinfo{author}{\bibfnamefont{F.}~\bibnamefont{Zhang}}
  (\bibinfo{year}{2014}), \eprint{arXiv:1406.4602}.

\bibitem[{\citenamefont{Yang et~al.}(2014)\citenamefont{Yang, Zimmerman, and
  Lehner}}]{Yang:2014tla}
\bibinfo{author}{\bibfnamefont{H.}~\bibnamefont{Yang}},
  \bibinfo{author}{\bibfnamefont{A.}~\bibnamefont{Zimmerman}},
  \bibnamefont{and} \bibinfo{author}{\bibfnamefont{L.}~\bibnamefont{Lehner}}
  (\bibinfo{year}{2014}), \eprint{arXiv:1402.4859}.

\bibitem[{\citenamefont{{Newman} and {Penrose}}(1962)}]{NewmanPenrose1962}
\bibinfo{author}{\bibfnamefont{E.}~\bibnamefont{{Newman}}} \bibnamefont{and}
  \bibinfo{author}{\bibfnamefont{R.}~\bibnamefont{{Penrose}}},
  \bibinfo{journal}{J. Math. Phys.} \textbf{\bibinfo{volume}{3}},
  \bibinfo{pages}{566} (\bibinfo{year}{1962}).

\bibitem[{\citenamefont{Stewart}(1991)}]{StewartBook}
\bibinfo{author}{\bibfnamefont{J.}~\bibnamefont{Stewart}},
  \emph{\bibinfo{title}{Advanced General Relativity}}
  (\bibinfo{publisher}{Cambridge University Press}, \bibinfo{year}{1991}).

\bibitem[{\citenamefont{Yang et~al.}(2013{\natexlab{a}})\citenamefont{Yang,
  Zhang, Zimmerman, Nichols, Berti et~al.}}]{Yang2012b}
\bibinfo{author}{\bibfnamefont{H.}~\bibnamefont{Yang}},
  \bibinfo{author}{\bibfnamefont{F.}~\bibnamefont{Zhang}},
  \bibinfo{author}{\bibfnamefont{A.}~\bibnamefont{Zimmerman}},
  \bibinfo{author}{\bibfnamefont{D.~A.} \bibnamefont{Nichols}},
  \bibinfo{author}{\bibfnamefont{E.}~\bibnamefont{Berti}},
  \bibnamefont{et~al.}, \bibinfo{journal}{Phys.Rev.D}
  \textbf{\bibinfo{volume}{87}}, \bibinfo{pages}{041502}
  (\bibinfo{year}{2013}{\natexlab{a}}), \eprint{1212.3271}.

\bibitem[{\citenamefont{Yang et~al.}(2013{\natexlab{b}})\citenamefont{Yang,
  Zimmerman, Zengino{\u{g}}lu, Zhang, Berti et~al.}}]{Yang:2013uba}
\bibinfo{author}{\bibfnamefont{H.}~\bibnamefont{Yang}},
  \bibinfo{author}{\bibfnamefont{A.}~\bibnamefont{Zimmerman}},
  \bibinfo{author}{\bibfnamefont{A.}~\bibnamefont{Zengino{\u{g}}lu}},
  \bibinfo{author}{\bibfnamefont{F.}~\bibnamefont{Zhang}},
  \bibinfo{author}{\bibfnamefont{E.}~\bibnamefont{Berti}},
  \bibnamefont{et~al.}, \bibinfo{journal}{Phys.Rev.}
  \textbf{\bibinfo{volume}{D88}}, \bibinfo{pages}{044047}
  (\bibinfo{year}{2013}{\natexlab{b}}), \eprint{1307.8086}.

\bibitem[{\citenamefont{{Detweiler}}(1980)}]{Detweiler1980}
\bibinfo{author}{\bibfnamefont{S.}~\bibnamefont{{Detweiler}}},
  \bibinfo{journal}{\apj} \textbf{\bibinfo{volume}{239}}, \bibinfo{pages}{292}
  (\bibinfo{year}{1980}).

\bibitem[{\citenamefont{{Hod}}(2008)}]{Hod2008a}
\bibinfo{author}{\bibfnamefont{S.}~\bibnamefont{{Hod}}},
  \bibinfo{journal}{\prd} \textbf{\bibinfo{volume}{78}}, \bibinfo{eid}{084035}
  (\bibinfo{year}{2008}), \eprint{0811.3806}.

\bibitem[{\citenamefont{Zimmerman and Mark}(2014)}]{ZimmermanDF}
\bibinfo{author}{\bibfnamefont{A.}~\bibnamefont{Zimmerman}} \bibnamefont{and}
  \bibinfo{author}{\bibfnamefont{Z.}~\bibnamefont{Mark}}
  (\bibinfo{year}{2014}), \bibinfo{note}{unpublished}.

\bibitem[{\citenamefont{Onozawa et~al.}(1996)\citenamefont{Onozawa, Mishima,
  Okamura, and Ishihara}}]{Onozawa:1995vu}
\bibinfo{author}{\bibfnamefont{H.}~\bibnamefont{Onozawa}},
  \bibinfo{author}{\bibfnamefont{T.}~\bibnamefont{Mishima}},
  \bibinfo{author}{\bibfnamefont{T.}~\bibnamefont{Okamura}}, \bibnamefont{and}
  \bibinfo{author}{\bibfnamefont{H.}~\bibnamefont{Ishihara}},
  \bibinfo{journal}{Phys.Rev.} \textbf{\bibinfo{volume}{D53}},
  \bibinfo{pages}{7033} (\bibinfo{year}{1996}), \eprint{gr-qc/9603021}.

\bibitem[{\citenamefont{Andersson and Onozawa}(1996)}]{Andersson1996}
\bibinfo{author}{\bibfnamefont{N.}~\bibnamefont{Andersson}} \bibnamefont{and}
  \bibinfo{author}{\bibfnamefont{H.}~\bibnamefont{Onozawa}},
  \bibinfo{journal}{Rev. D} p. \bibinfo{pages}{9607054} (\bibinfo{year}{1996}).

\bibitem[{\citenamefont{{Mashhoon}}(1985)}]{Mashhoon1985}
\bibinfo{author}{\bibfnamefont{B.}~\bibnamefont{{Mashhoon}}},
  \bibinfo{journal}{\prd} \textbf{\bibinfo{volume}{31}}, \bibinfo{pages}{290}
  (\bibinfo{year}{1985}).

\bibitem[{\citenamefont{Yang et~al.}(2012)\citenamefont{Yang, Nichols, Zhang,
  Zimmerman, Zhang et~al.}}]{Yang2012a}
\bibinfo{author}{\bibfnamefont{H.}~\bibnamefont{Yang}},
  \bibinfo{author}{\bibfnamefont{D.~A.} \bibnamefont{Nichols}},
  \bibinfo{author}{\bibfnamefont{F.}~\bibnamefont{Zhang}},
  \bibinfo{author}{\bibfnamefont{A.}~\bibnamefont{Zimmerman}},
  \bibinfo{author}{\bibfnamefont{Z.}~\bibnamefont{Zhang}},
  \bibnamefont{et~al.}, \bibinfo{journal}{Phys.Rev.}
  \textbf{\bibinfo{volume}{D86}}, \bibinfo{pages}{104006}
  (\bibinfo{year}{2012}), \eprint{1207.4253}.

\bibitem[{\citenamefont{Geroch et~al.}(1973)\citenamefont{Geroch, Held, and
  Penrose}}]{GHP}
\bibinfo{author}{\bibfnamefont{R.}~\bibnamefont{Geroch}},
  \bibinfo{author}{\bibfnamefont{A.}~\bibnamefont{Held}}, \bibnamefont{and}
  \bibinfo{author}{\bibfnamefont{R.}~\bibnamefont{Penrose}},
  \bibinfo{journal}{Journal of Mathematical Physics}
  \textbf{\bibinfo{volume}{14}}, \bibinfo{pages}{874} (\bibinfo{year}{1973}),
  \urlprefix\url{http://scitation.aip.org/content/aip/journal/jmp/14/7/10.1063/1.1666410}.

\end{thebibliography}

\end{document}